\documentclass{elsart}

\usepackage{graphicx}
\usepackage{epsfig}
\usepackage{natbib}
\usepackage{amssymb}

\begin{document}

\begin{frontmatter}



\title{Photodissociation of small carbonaceous molecules of astrophysical
interest}


\author[label1]{M.C. van Hemert} 
\author[label2]{E.F. van Dishoeck\corauthref{cor1}}

\corauth[cor1]{Corresponding author}
\ead{ewine@strw.leidenuniv.nl}

\address[label1]{Leiden Institute of Chemistry, Gorlaeus Laboratories,
Leiden University, P.O. Box 9502, 2300 RA Leiden, The Netherlands}

\address[label2]{Leiden Observatory, Leiden University, P.O.\ Box 9513,
2300 RA Leiden, The Netherlands}

\begin{abstract}

Astronomical observations have shown that small carbonaceous molecules
can persist in interstellar clouds exposed to intense ultraviolet
radiation.  Current astrochemical models lack quantitative information
on photodissociation rates in order to interpret these data. We here
present {\it ab initio} multi-reference configuration-interaction
calculations of the vertical excitation energies, transition dipole
moments and oscillator strengths for a number of astrophysically
relevant molecules: C$_3$, C$_4$, C$_2$H, $l-$ and $c-$C$_3$H, $l-$
and $c-$C$_3$H$_2$, HC$_3$H, $l-$C$_4$H and $l-$C$_5$H. Highly excited
states up to the 9'th root of each symmetry are computed, and several
new states with large oscillator strengths are found below the
ionization potentials.  These data are used to calculate upper limits
on photodissociation rates in the unattenuated interstellar radiation
field by assuming that all absorptions above the dissociation limit
lead to dissociation.

\end{abstract}

\end{frontmatter}

\section{Introduction}
\label{intro}
Of the more than 130 different molecules found in interstellar space,
an important class is formed by the unsaturated carbonaceous
species. Carbohydrides ranging from small molecules such as C$_2$H
\cite{Tucker74} to long chains like C$_8$H \cite{Bell99} and
HC$_{11}$N \cite{Bell97} have been detected through their millimeter
transitions in cold dark clouds like TMC-1 for several decades. Some
of these molecules have also been seen in their cyclic form, with
c-C$_3$H$_2$ as the best-known example \cite{Thaddeus85b}. Even
negative ions, in particular C$_6$H$^-$, have now been detected
\cite{McCarthy06}. Bare carbon chains are likely present as well but
do not have a permanent dipole moment and can therefore not be
observed through their pure rotational transitions in
emission. Instead, the smallest members of this family, C$_2$ and
C$_3$, have been detected in diffuse interstellar clouds, --i.e.,
clouds which are not completely opaque to visible and ultraviolet
radiation-- through their electronic absorptions against bright
background stars \cite{Hobbs79,vanDis84,Maier01}. $c-$C$_3$H$_2$ has
been seen in diffuse clouds as well, through absorption at radio
wavelengths against background quasars \cite{Cox88}. The relatively
large abundances of these non-saturated molecules, in spite of the
fact that there is $10^4$ times more hydrogen than carbon in
interstellar clouds, are a vivid demonstration that interstellar
chemistry is not in thermodynamic equilibrium. Instead, the kinetics
of the reactions that form and destroy these molecules need to be
taken into account explicitly in order to explain their abundances. In
cold dark clouds, these traditionally involve a series of ion-molecule
and neutral-neutral reactions \cite{Winnewisser93,Herbst03}.

More recently, several of the carbohydrides C$_n$H$_m$ have also been
observed in so-called Photon-Dominated Regions
(PDRs) \cite{Hollenbach97}, i.e., clouds which are exposed to intense
ultraviolet (UV) radiation. Species like C$_2$H and c-C$_3$H$_2$ have
been detected in the Orion Bar PDR \cite{Hogerheijde95} and even more
complex molecules like C$_3$H$_2$ and C$_4$H have been seen in the
Horsehead nebula, M38 \cite{Teyssier04,Pety05}.  Together
with the above mentioned observations of diffuse clouds, the data
demonstrate that these molecules can exist in UV-exposed environments
with abundances comparable to those in cold dark clouds.  Traditional
PDR chemical models cannot explain the high abundances of these
molecules, leading to speculations that they are perhaps produced by
fragmentation of even larger carbonaceous molecules such as polycyclic
aromatic hydrocarbons (PAHs) \cite{Pety05}. However, a major
uncertainty in these and other models are the photodissociation rates
of the carbohydrides, which are basically unknown.

Other regions in which ultraviolet photons play a role in the
chemistry include the envelopes of evolved stars and cometary
atmospheres. The same carbohydride molecules have been detected
in the envelope of the carbon-rich late-type star IRC+10216, where
their similar distributions are a puzzle for photochemical
models \cite{Guelin00}.  In cometary atmospheres, species as complex as
C$_4$H have been inferred, whose presence is not readily explained by
the standard parent-daughter photochemical models \cite{Geiss99}.

Many theoretical studies of the electronic structure of small
carbohydride molecules exist in the literature
\cite[e.g.,]{Lee85,Botsch89,Jonas92,Cui97,Carter00,Orden98}. However,
most of them are limited to the ground and lowest few excited
electronic states. Only few studies of higher states exist, most of
them performed by Peyerimhoff and her co-workers starting more than 30
years ago, e.g., \cite{Shih77,Romelt78} As will be shown here, those
results are still highly relevant and provide an excellent starting
point for further studies.

The aim of this paper is to provide insight into the photodissociation
processes of carbohydride molecules through {\it ab initio} quantum
chemical calculations of the vertical excitation energies and
oscillator strengths for as large a number of excited electronic
states as feasible with current programs.  By assuming that most of
the absorptions with energies above the dissociation limit lead to
destruction, estimates of the (upper limits of) photodissociation
rates under interstellar conditions can be obtained. Even though such
calculations are necessarily limited to the smaller members of the
carbo-hydride family, they do provide quantitative constraints to test
the basic interstellar chemical networks. Specifically, C$_2$H, C$_3$,
C$_4$, $l-$ and $c-$C$_3$H, $l-$ and $c-$ C$_3$H$_2$, HC$_3$H,
$l-$C$_4$H and $l-$C$_5$H are studied here. Except for HC$_3$H, all
these species have been chosen to have been observed in interstellar
space.

\section{Methods} 

\subsection{Electronic structure calculations}

For small diatomic and triatomic molecules, quantum chemical
calculations of the potential energy surfaces and transition dipole
moments combined with dynamical calculations of the nuclear motions
can provide photodissociation cross sections and oscillator strengths
that agree with experiments to better than 20--30\%, e.g.,
\cite{vanDis82,vanHar01,Schinke93} (see \cite{Kirby88} for an early
review).  For the polyatomic molecules considered here, calculations
of the full potential surfaces including all degrees of freedom become
very time consuming, as do the multi-dimensional dynamics. Moreover,
such detail is not needed in order to compute photorates, since those
are largely determined by the potentials and transition moments in the
Franck-Condon region.  The simplest alternative is therefore to only
compute the vertical excitation energies and transition dipole moment
of the molecule at its equilibrium position and assume a certain
probability that absorption into each excited state above the
dissociation limit leads to dissociation. For the examples cited
above, such an approach leads to similar rates within the accuracy of
the calculations.  The focus of our calculations is therefore on
electric dipole-allowed transitions to states lying above the lowest
dissociation limit but below 13.6 eV. The interstellar radiation field
has a broad spectrum from visible to extreme ultraviolet wavelengths,
peaking in intensity around 7 eV and cutting off at the atomic H
ionization potential at 13.6 eV \cite{Draine78}.

All calculations presented here were performed with the Wuppertal-Bonn
MRDCI set of programs as implemented in the GAMESS-UK program package
version 7.0 \cite{GAMESS}. For C, the TZVP atomic orbital basis set was
used, and for H, the DZP basis set \cite{Dunning71}.  To allow for a
proper description of molecular Rydberg states, two diffuse $p$ and
two diffuse $d$ functions were put on specific sites. For C$_2$ and
C$_2$H, this was the middle of the CC bond; for C$_2$ and C$_4$H the
middle of the C1-C2 and C3-C4 bonds; for C$_3$, $l-$C$_3$H,
$l-$C$_3$H$_2$ and HC$_3$H the middle C; for C$_5$ on C2 and C4; for
$c-$C$_3$H on the C connected to H; and for $c-$C$_3$H$_2$ on the lone
C. These sites were chosen to avoid linear dependencies in the atomic
orbital basis set due to large overlap of diffuse functions positioned
on adjacent atoms.  Cartesian $d$ functions were used, of which the
spherical component served as the $3s$ basis function.

Molecular orbitals were generated using up to the maximum number of
reference states, 255.  The selection threshold was generally set at
0.5 $\mu$Hartree.  The total number of configurations included in the
configuration interaction (CI) calculation ranged from 200000--400000
per symmetry. The sum over the coefficients-squared in the final CI
wave function is typically 0.95 for the smaller species, dropping to
0.9 for the larger molecules. The CI energies were extrapolated and
corrected using the Davidson extrapolation \cite{Davidson}. The aim
was to compute as many excited electronic states as feasible, up to 9
per symmetry. For the lower states, comparison with existing
calculations and experiments indicates accuracies within 0.3 eV,
generally better. Oscillator strengths to the lower states agree
within 30\% or better. For the higher states, typically the 5'th root
and higher per symmetry, the accuracy decreases because many states
and orbitals can mix. Nevertheless, such calculations should still
provide insight into the location of those states, in particular
whether they are above or below the ionization potential and below the
13.6 eV cutoff of the interstellar radiation field. Moreover, the
magnitude of the oscillator strengths (strong or weak) should be
reliable. Note that the precise values of the excitation energies are
not so important for the purposes of calculating interstellar
photorates because of the broad range of incident energies. The only
exception is possible overlap with Lyman $\alpha$ radiation at 10.2
eV, which is important for certain astrophysical environments
\cite{vanDis06}.

In all calculations, the largest Abelian subgroup of the full
$C_{\infty v}$ group, $C_{2v}$, was used, with the molecule put along
the $z-$axis. The doubly degenerate states are then split into the
$B_1$ and $B_2$ irreducible representations for $\Pi$ and $\Phi$
states, and into $A_1$ and $A_2$ for the $\Delta$ states. In general,
the degeneracies for the $\Delta$ states are recovered within a few
hundreds of an eV in the calculations. The average of the two $A_1$
and $A_2$ values is tabulated here. The $A_1$ irreducible
representation also contains the $\Sigma^+$ states, and the $A_2$ the
$\Sigma^-$ states.  Some excited states with electronic angular
momentum higher than 2 (in particular $\Phi$ states) are also found in
our calculations but they are not tabulated here since the electric
dipole transition moments from the ground state to these states are
zero, implying that they do not contribute to photodissociation.  The
valence or Rydberg character of the states is determined from
calculations of the $x^2$, $y^2$ and $z^2$ expectation values.

Calculations have been limited to the equilibrium geometry
corresponding to the lowest energy. For C$_3$, C$_2$H, C$_4$, C$_4$H
and C$_5$H this is the linear geometry.  C$_3$H and C$_3$H$_2$ have
been detected in various isomeric forms so both cyclic and (near-)linear
forms have been studied.

The dissociation energies for C$_n$H are calculated as the difference
between the ground state energy of C$_n$H at its equilibrium geometry
and the energy of ground state C$_n$, also at its equilibrium
geometry, calculated in the presence of an H atom positioned at a
distance of 20 Bohr from the center of mass.  In this way the errors
associated with the lack of size consistency are significantly
reduced. Note that the spatial symmetry of the molecule and the
fragment generally differs.

\subsection{Photodissociation rates}

Oscillator strengths to all excited states have been calculated
according to
\begin{equation}
f_{ul}^{el} = {2\over 3}\ g_{ul} \ \Delta E_{ul}\  \mu_{ul}^2
\end{equation}
where all quantities are in atomic units (a.u.), $\mu_{ul}$ is the
transition dipole moment from lower state $l$ to upper state $u$, and
$g_{ul}$ is a degeneracy factor which is 2 for a $\Pi \leftarrow \Sigma $
transitions and 1 for any other transition.  For linear molecules, the
dipole moment operator contained in $\mu_{ul}$ corresponds to $-\sum
z_j$ for transitions between states with the same value of the
electronic angular momentum projection quantum number $\Lambda$ and to
$-\sum (x_j+iy_j)/\sqrt{2}$ for transitions with $\Delta \Lambda = \pm
1$. The computed excitation energies $\Delta E_{ul}$ were used in this
formula, not any experimental values. Hence, differences with other
work can stem from differences in both $\mu_{ul}$ and in $\Delta
E_{ul}$.

The photodissociation rate of a molecule can be computed from
\begin{equation}
k_{\rm pd}^{\rm cont} = \int \sigma(\lambda) I(\lambda) d\lambda \ \ \
{\rm s^{-1}}
\end{equation}
where $\sigma$ is the photodissociation cross section in cm$^2$ and
$I$ is the mean intensity of the radiation in photons cm$^{-2}$
s$^{-1}$ \AA$^{-1}$ \cite{vanDis06}. Under interstellar conditions,
only single photon processes are important.  For photodissociation
initiated by line absorptions (e.g., predissociation), the rate
becomes
\begin{equation}
k_{\rm pd}^{\rm line}= {{\pi e^2} \over m c^2} \lambda_{ul}^2 f_{ul} \eta_u
 I_{ul}
 \ \ \ {\rm s^{-1}} 
\end{equation}
where $\eta_u$ the dissociation efficiency of state $u$, which lies
between 0 and 1. The numerical value of the factor $\pi e^2/mc^2$ is
$8.85 \times 10^{-21}$ in the adopted units with $\lambda$ in \AA. The
total photodissociation rate of a molecule is obtained by summing over
all channels.  In this work, no dynamical calculations are performed
to obtain continuous cross sections for dissociative states.  Hence,
all photorates are computed with Eq.\ (3). It is furthermore assumed
that the dissociation efficiency $\eta_u$=1, either through direct
dissociation in a repulsive state or by predissociation. The
motivation for this choice is that for larger molecules, internal
conversion to a lower (dissociative) electronic state is usually much
more rapid than any radiative decay rates owing to the high density of
states \cite{Ashfold87}.  Specific experimental evidence of high
dissociation efficiencies will be presented in the results section for
individual molecules.

\section{Results} 

\subsection{C$_2$H}

The C$_2$H radical, detected in interstellar clouds since 1974
\cite{Tucker74}, is an important step in the formation of longer
carbon chains. In comets, it could be a photodissociation product of
C$_2$H$_2$ and a precursor of the widely observed C$_2$
molecule. Given also its importance in combustion processes, this
radical has received ample theoretical attention, starting with the
papers by Shih, Peyerimhoff and co-workers \cite{Shih77,Shih79} and
culminating with the more recent two-dimensional potential surfaces by
Duflot et al. \cite{Duflot94}.

As a test of our computational procedure, we present in
Table~\ref{tabc2h} our computed vertical excitation energies,
oscillator strengths and the main configurations of each state. Our
energies generally agree well within 0.3 eV with those of Koures and
Harding \cite{Koures91} (their CI + DV2 results) and lie in between
those of Duflot et al. \cite{Duflot94} and Shih et al. \cite{Shih77}.
Close inspection shows that the larger differences are usually for
states with Rydberg character (which are not included in Duflot et
al. \cite{Duflot94}) or for states with a strong interaction with a
neighboring state, leading sometimes even to a switch in the character
of the state.  For example, the 4$^2\Sigma^+$ state in our calculation
has Rydberg $3p_\pi$ character whereas it has ... $4\sigma^1 1\pi^3
5\sigma^2 3\pi^1$ in Duflot et al., resulting also in a very different
transition dipole moment (see below).

Our computed $A^2\Pi - X ^2\Sigma^+$ transition dipole moment of 0.22
a.u.\ is very close to that of 0.23 a.u.\ computed by Duflot et
al. \cite{Duflot94} and Peric et al. \cite{Peric90}. Because of the
slightly higher excitation energy in our work, the oscillator
strengths show somewhat larger differences. Comparison for other
states is difficult since Duflot et al.\ do not give any numerical
values. However, their Figure 7 shows that neither the 3 and 4
$^2\Sigma^+$, nor the 2 and 3 $^2\Pi$ states have significant
transition dipole moment. This is generally consistent with our
results. For the 5$^2\Sigma^+$ state we find a huge
transition dipole moment of 1.1 a.u.\ due to the Rydberg
character of our wavefunction. Thus, this state around 10 eV will
dominate the interstellar C$_2$H photodissociation but the higher
$^2\Pi$ states in the 8.5--10.5 eV range can also contribute
significantly up to the ionization potential of $\sim$11.4 eV
(Table~\ref{tabdiss}).

\subsection{C$_3$}

The C$_3$ molecule was detected in cometary spectra in 1882 through
its $A-X$ system at 4050 \AA \ \cite{Huggins82} and in interstellar
clouds more than a century later \cite{Haffner95,Maier01}.  The
molecule is also seen in the atmospheres of cool carbon stars through
its mid-infrared \cite{Hinkle88} and far-infrared \cite{Cerni00}
transitions (see \cite{Orden98} for review).

C$_3$ is a linear molecule with a ground X$^1\Sigma_g^+$ state.  Its
dissociation energy is computed to be about 4.6 eV, whereas its
ionization energy is around 12 eV (Table~\ref{tabdiss}).  Quantum
chemical studies range from the early work by Chabalowski et al.\
\cite{Chaba86} to the recent calculations by Terentyev et
al. \cite{Terent04}.  Table~\ref{tabc3} summarizes our computed
excitation energies, together with the oscillator strengths and main
configurations. For the low-lying valence states, our results agree to
better than 0.3 eV with the MR-AQCC values of Monninger et al.\
\cite{Monninger02}.  These results also show that the $^1\Sigma_u^+
\leftarrow X^1\Sigma_g^+$ ($1\pi_u \to 1\pi_g$) transition at 8.17 eV
has by far the largest oscillator strength, as predicted first by
Pitzer and Clementi \cite{Pitzer59}.  This is confirmed by the
combined experimental and theoretical study by Monninger et al.\
\cite{Monninger02} whose 1100--5600 \AA \ spectrum demonstrates that
the $^1\Sigma_u^+ \leftarrow X^1\Sigma_g^+$ transition around 1700 \AA
\ is indeed the strongest band.  Their experiments in Ne and Ar
matrices show a broad band but with some progressions superposed,
which are evidence for interstate vibronic coupling with adjacent
$\Pi_g$ states. Indeed, the $^1\Sigma_u^+$ state is predicted to be
unstable to bending, leading to avoided crossings with the lower-lying
$\Pi_g$ states \cite{Monninger02}. These interactions can also lead to
dissociative channels to produce C$_2$ + C.

\subsection{C$_3$H}

Both linear and cyclic C$_3$H have been detected in the interstellar
medium by their transitions at millimeter wavelengths
\cite{Thaddeus85a,Yama87}. The $^2\Pi$ ground state of $l-$C$_3$H lies
about 0.4--0.6 eV above the $^2B_2$ ground state of $c-$C$_3$H
\cite{Takahashi96}, with a small barrier toward the cyclic
state. Hence, when C$_3$H is produced in its linear form by some
sequence of ion molecule or other chemical processes, it can be
stable, and both isomers are therefore considered in this work.

\subsubsection{$l-$C$_3$H}

The geometry of the $1 ^2\Pi$ state was taken from the experimental
work of McCarthy and Thaddeus \cite{McCarthy05}.  Table~\ref{tabl-c3h}
summarizes our computed vertical excitation energies, oscillator
strengths, and the corresponding configurations.  Comparison with the
CASSCF results of Ding et al.\ \cite{Ding01} shows good agreement for
the lowest states. The dissociation energy of $l-$C$_3$H to C$_3$ + H
is computed at 3.3 eV. Of the dipole-allowed states above this
dissociation limit, the higher $^2\Pi$ states around 7.8 eV have the
largest oscillator strengths, in particular the 8$^2\Pi$ state, which
will dominate the photodissociation of the molecule.

\subsubsection{$c-$C$_3$H}

The $c-$C$_3$H radical has $C_{2v}$ geometry in its $1 ^2B_2$ ground
state, with the equilibrium structure for our calculations taken from
Yamamoto and Saito \cite{Yama94}.  Table~\ref{tabc-c3h} summarizes our
computed vertical excitation energies, oscillator strengths and
configurations.  Comparison with the CASSCF results of Ding et al.\
\cite{Ding01} shows again good agreement for the lowest states to
0.1--0.2 eV. The dissociation energy of $c-$C$_3$H to C$_3$ + H is
around 4.3 eV whereas its ionization potential is computed to lie at
9.6 eV (Table~\ref{tabdiss}). Of the dipole-allowed states above the
dissociation limit, the $2 ^2A_1$ state around 5 eV and the higher
$^2A_2$ states around 7.5 eV have the largest oscillator strengths,
but the sum over the other states is comparable.  Thus, there appear
to be many potential routes to photodissociation for $c-$C$_3$H.

\subsection{C$_3$H$_2$}

Cyclopropenylidene, $c-$C$_3$H$_2$, was the first cyclic molecule to
be detected in the interstellar medium \cite{Thaddeus85b} and
subsequently found to be ubiquitous throughout the Galaxy, even in
diffuse gas \cite{Matthews85,Cox88}. The near-linear form HCCCH
(propargylene), denoted here as HC$_3$H, lies 0.4--0.8 eV higher in
energy \cite{Seburg97,Takahashi96}, depending whether the zero-point
vibrational energy is included. Since it has a near-zero dipole
moment, it has not yet been observed in interstellar clouds through
radio transitions but is likely present as well. There are various
other stable isomers of C$_3$H$_2$, of which another linear form,
H$_2$CCC (vinylidenecarbene or propadienylidene), denoted here as
$l-$C$_3$H$_2$, was discovered in interstellar space in 1991
\cite{Cerni91}. It lies $\sim$0.6 eV above the cyclic ground state
including zero-point vibrational energy corrections
\cite{Seburg97}. Hence, all three isomers are considered in this
work. Although there are many theoretical studies of the ground states
of the various C$_3$H$_2$ isomers dating back to 1976
\cite{Hehre76,Lee85,DeFrees86}, the excited states are largely
unexplored.

The equilibrium structures of the electronic ground states are taken
from Seburg et al., their Figure~5 \cite{Seburg97}. Our calculated
relative ordering of the C$_3$H$_2$ isomers is consistent with
previous findings but due to different amounts of recovered
correlation energy, our energy splittings are larger than in other
studies.

The adiabatic ionization potentials of the various C$_3$H$_2$ isomers
lie at 8.96 (HC$_3$H), 9.2 ($c-$C$_3$H$_2$) and 10.4 eV
($l-$C$_3$H$_2$), respectively (Table~\ref{tabdiss}). Above these
thresholds, the photoionization efficiency of all three C$_3$H$_2$
isomers increases rapidly so that photoionization will become the main
pathway \cite{Taatjes05}. Some of these ionizations are likely to be
dissociative, but this option is not considered here.

\subsubsection{HC$_3$H}

The HC$_3$H radical has $C_2$ geometry with a $^3B$ ground state.  In
this case, the $C_2$ axis was put through the middle C atom and the
middle of the H1--H2 line. Table~\ref{tabl-c3h2} summarizes our
computed vertical excitation energies, oscillator strengths and
configurations.  For the lowest four excited states, our results are
in good agreement with those of Mebel et al.\ \cite{Mebel98} (their
MRCI+D(4,8)/ANO(2+) results). The dissociation energy of HC$_3$H to
$l-$C$_3$H + H is computed to be around 3.1 eV, with the dissociation
energy to C$_3$ + H$_2$ perhaps even less \cite{Mebel98}.
Of the dipole-allowed states above the dissociation and below the
ionization limit at 8.96 eV, the higher $^3A$ states at 6.2 and 7.5
eV, both of which have $\pi \to \pi^*$ character, have the largest
oscillator strengths.

\subsubsection{$c-$C$_3$H$_2$}

The $c-$C$_3$H$_2$ molecule has a $^1A_1$ ground state in its $C_{2v}$
geometry, with a large dipole moment of 3.4 Debye owing to the two
unpaired electrons on one of the three carbon atoms.
Table~\ref{tabc-c3h2} summarizes our computed vertical excitation
energies and oscillator strengths, which again agree well in terms of
energies with those of Mebel et al.\ \cite{Mebel98}, although our
oscillator strengths are somewhat lower. Compared with $c-$C$_3$H,
$c-$C$_3$H$_2$ has only few low-lying electronic states, and all of
the dipole-allowed transitions lie above the dissociation energy of
4.4 eV \cite{Mebel98}. Those to the $6 ^1A_1$ ($\pi \to \pi^*$) at
9.4, $7 ^1B_1$ at 10.6 and $9 ^1B_1$ state at 11.2 eV have the largest
oscillator strengths. However, none of these states lie below the
ionization potential of $c-$C$_3$H$_2$ at 9.2 eV. Thus, its
photodissociation rate will differ substantially whether or not these
states are included (see \S 4).

\subsubsection{$l-$C$_3$H$_2$}

The vinylidenecarbene or propadienylidene isomer of C$_3$H$_2$,
denoted here as $l-$C$_3$H$_2$, also has $^1A_1$ symmetry in its
$C_{2v}$ ground state geometry.  Our computed vertical excitation
energies (Table~\ref{tabv-c3h2}) of 1.86, 2.44 and 5.54 eV to the
$\tilde A ^1A_2$, $\tilde B ^1B_1$ and $\tilde C (2) ^1A_1$ states are
consistent with the experimental adiabatic values \cite{Hodges00} of
1.73, 2.00 and 4.84 eV, respectively, and agree within 0.1--0.2 eV
with the vertical values computed by Mebel et al.\ \cite{Mebel98}
The dissociation energy of $l-$C$_3$H$_2$ to C$_3$H + H is computed to
lie around 3.9 eV, so that only the $\tilde C (2) ^1A_1$  and
higher states can lead to photodissociation. The $\tilde C$ state shows a
well-resolved progression in its electronic spectrum \cite{Hodges00},
but the resolution of those data is not high enough to measure
predissociation rates. Thus, we consider the photodissociation both
with and without taking the $\tilde C$ state into account.

Of the dipole-allowed transitions below the ionization potential at
10.4 eV, those to the higher $ ^1A_1$ states around 9 eV have the
strongest oscillator strengths. These states all have $2b_2 \to n^*b_2$
character and likely belong to the Rydberg series converging to the
lowest ionization potential. Similarly, the higher $^1B_2$ states
belong to the $2b_2 \to na_1$ Rydberg series.

\subsection{C$_4$}

The lowest energy $X ^3\Sigma_g^-$ ground state of C$_4$ occurs for
the linear geometry. C$_4$ has been searched for in diffuse
interstellar clouds through its $^3\Sigma_u^- \leftarrow ^3\Sigma_g^-$
transition around 3789 \AA \ but not yet detected
\cite{Maier02}. However, a pattern of bands at 57 $\mu$m observed with
the {\it Infrared Space Observatory} toward a handful of objects
\cite{Cerni02} is consistent with transitions in the $\nu_5$ bending
mode \cite{Moazzen94}.  Rhombic C$_4$, which is almost isoenergetic
with $l-$C$_4$, has not yet been detected in interstellar space.

Our calculations use the $l-$C$_4$ equilibrium values computed by
Botschwina \cite{Botsch97}, which are consistent with experiments
\cite{Orden98}.  The results are presented in Table~\ref{tabc4}.  Our
excitation energies of the lower states are in good agreement with
experiments and with the calculations of Mass\'o et al.\
\cite{Masso06}, except for the 2$^3\Delta_u$ state.  Note that the
experimental values refer to $T_0$ rather than $T_e$, which can differ
by a few tenths of eV.  Our computed dissociation energy $D_e$ is 4.7
eV, in good agreement with the experimentally inferred $D_0$ value of
4.71 $\pm$0.15 eV \cite{Choi00}.

The lowest excited state above the dissociation limit to which dipole
allowed transitions are possible is the $3 ^3\Pi_u$ state at 5.1 eV.
Indeed, photofragment yield spectra in the 2.22--5.40 eV range by Choi
et al.\ \cite{Choi00} show significant single-photon dissociation
into C$_3$ + C at $\geq$5.2 eV, with a minor channel to C$_2$ +
C$_2$. Their spectra are well reproduced by phase space theory models in
which the product state distributions are statistical. This implies
that absorption into the excited state is most likely followed by rapid
internal conversion to the ground state potential energy surface with
no barriers present along the dissociation coordinates.

By far the strongest absorption occurs into the $2^3\Sigma_u^-$ state
around 6.95 eV, which has an oscillator strength of 1.56. Even though
higher $\Sigma_u^-$ and $\Pi_u$ states will contribute, the
$2^3\Sigma_u^-$ channel will dominate the interstellar
photodissociation of C$_4$, if indeed every absorption is followed by
dissociation.

\subsection{$l-$C$_4$H}

The $l-$C$_4$H (butadiynyl) radical was detected in the envelopes of
carbon-rich evolved stars nearly 30 years ago \cite{Guelin78}. It was
subsequently found to be very abundant in cold dark clouds like
TMC-1 \cite{Irvine81} and even detected in comets \cite{Geiss99}. It is one
of the carbon-bearing molecules found at the edges of
PDRs \cite{Pety05}, where it can photodissociate into smaller species.

C$_4$H has linear symmetry with a $^2\Sigma^+$ ground state. The
equilibrium coordinates in our calculations were taken from
Ref. \cite{McCarthy05}. Table~\ref{tabc4h} summarizes our computed
vertical excitation energies, oscillator strengths and configurations.
Comparison with the results of Graf et al.\ \cite{Graf01} shows good
agreement in both energies and transition dipole moments for the
lowest states, but poor agreement for the higher $^2\Pi$ states, where
the Graf et al.\ energies are generally lower by up to 1 eV.  There
are two main resons for this. First, Graf et al.\ did not include
diffuse (Rydberg) functions in their basis set, which start to become
important for the higher states. Second, their CASPT2 perturbation
method does not guarantee a lower bound to the energies.  Our
calculations agree in the fact that none of the higher lying $^2\Pi$
states have large oscillator strengths.

The threshold for photodissociation, corresponding to dissociation of
C$_4$H to C$_4$ + H, lies around 4.7 eV (Table~\ref{tabdiss}).  Of the
dipole-allowed transitions below the ionization potential at 9.6 eV,
the 4 and 5 $^2\Sigma^+$ states at 7.7 and 8.7 eV, respectively, have
orders of magnitude larger oscillator strengths than other states and
will thus dominate the photodissociation. Both states have $1\pi \to
n\pi$ Rydberg character. Thus, our overall photodissociation rate will
be significantly larger than that using the data from Graf et
al.\ \cite{Graf01}

\subsection{$l-$C$_5$H}

Like C$_4$H, $l-$C$_5$H is detected toward the carbon-rich evolved
star IRC+10216 \cite{Cerni86} and in cold dark clouds \cite{Cerni87}.
It also has linear symmetry but with a $^2\Pi$ ground state. The
equilibrium coordinates in our calculations were taken from Ref.\
\cite{McCarthy05}. Table~\ref{tabc5h} summarizes our
results. Comparison with Haubrich et al.\ \cite{Haubrich02} shows
excellent agreement in both energies and oscillator strengths, except
for the higher-lying $^2\Sigma^+$ states. This difference is likely
due to the explicit inclusion of Rydberg states in our work. For the 3
and 4 $^2\Pi$ states, the oscillator strengths differ by a factor of 2
but these states show considerable interaction between the $2\pi \to
3\pi$ and $3\pi \to 4\pi$ $(\pi \to \pi^*)$ excitations. Differences
in mixing ratios can lead to large changes in transition dipole
moments to individual states, but not in energies. The computed
excitation energies for the lowest two excited states also agree well
with those measured and computed by Ding et al.\ \cite{Ding02}.

The threshold for photodissociation lies around 3.6 eV
(Table~\ref{tabdiss}).  Of the dipole-allowed transitions below the
ionization potential at 7.4 eV, both the 4 $^2\Pi$ ($2\pi \to 3\pi$)
state at 4.2 eV and the 6 $^2\Pi$ ($1\pi \to 3\pi)$ state around 6.1
eV have large oscillator strengths. All other states have typical
oscillator strengths of a few $\times 10^{-3}$ and thus contribute at
a lower level.

\section{Interstellar photodissociation rates}

In Table~\ref{tabisrates}, the computed photodissociation rates in the
unshielded interstellar radiation field cf.\ Draine \cite{Draine78}
are presented, using the oscillator strengths given in
Tables~\ref{tabc2h} to \ref{tabc5h} and asuming $\eta_u$=1 for all
states. Thus, these rates should be regarded as upper limits. Only
states above the dissociation limit and below the ionization potential
have been taken into account (see Table~\ref{tabdiss} for adopted
values). No corrections have been made for possible higher-lying
states below the ionization limit not computed in this work. Since the
oscillator strengths for Rydberg states decrease roughly as $1/n^3$,
it is assumed that any such corrections would be small since the
lowest Rydberg members are calculated explicitly. For reference,
inclusion of a hypothetical state at 9 eV with an oscillator strength
of 0.1 would increase the photodissociation rates by only $3.5\times
10^{-10}$ s$^{-1}$.

For C$_2$H, our new rate is a factor of 3 larger than that given in
van Dishoeck et al. \cite{vanDis06}, which was based on the energies
and oscillator strengths of Shih et al. \cite{Shih77}. The increase is
mostly due to the higher $^2\Sigma^+$ and $^2\Pi$ states which were
not computed in that work. For C$_3$, the new rate is only 30\%
larger, mostly because the photodissociation rate of this molecule is
dominated by the very strong absorption into the $1 ^1\Sigma_u^+$
state around 8 eV, which was included in previous estimates.

It is seen that the rates for the various carbon-bearing molecules
span a range of a factor of 8, with the rates for the bare carbon
chains (C$_3$, C$_4$) being largest and those for the odd-numbered
C$_n$H species lowest.  Of the different C$_3$H$_2$ isomers,
$l-$C$_3$H$_2$ has the largest photodissociation rate and
$c-$C$_3$H$_2$ the smallest by a factor of 3. However, $c-$C$_3$H$_2$
differs from the other isomers in that it has several states with
large oscillator strengths above the ionization potential of 9.15
eV. If those states were included, the $c-$C$_3$H$_2$
photodissociation rate would be increased by a factor of 2.  For
$l-$C$_3$H$_2$, the rate would drop from $5.1\times 10^{-9}$ to
$4.1\times 10^{-9}$ s$^{-1}$ if the $\tilde C$ state is not included.

In spite of the range of values, all rates are above $10^{-9}$
s$^{-1}$, coresponding to a lifetime of less than 30 yr at the edge of
an interstellar cloud. In regions such as the Orion Bar and the
Horsehead nebula, the radiation field is enhanced by factors of
$10^3-10^5$ compared with the standard field adopted here, decreasing
the lifetimes to less than 1 month.  Thus, there must be rapid
production routes of these molecules in order to explain their high
abundances in UV-exposed regions. This conclusion is not changed if
only 10\%, say, of the absorptions would lead to dissociation rather
than the 100\% assumed here. As argued in \S 2 and 3, it is plausible
that a substantial fraction of the absorptions lead to dissociation
for these larger molecules so that the upper limits should be close to
the actual values.

\section{Conclusions}

We have presented vertical excitation energies, transition dipole
moments and oscillator strengths for states up to the 9'th root of
each symmetry for several carbonaceous molecules of astrophysical
interest.  For lower-lying states, good agreement is generally found
with previous studies.  Several new, higher-lying states with large
oscillator strengths, often of Rydberg character, are revealed in this
work.

The calculated photodissociation rates of the small carbon-bearing
molecules studied here are substantial, leading to lifetimes at the
edges of interstellar clouds of less than 30 yr. These high rates
assume that all absorptions above the dissociation limit indeed lead to
dissociation, so that the rates should be viewed as upper
limits. Further experimental work is needed to quantify the
dissociation efficiencies for the strongest states found in this work.

\section{Acknowledgments}

Our calculations of the photodissociation processes of astrophysical
molecules, which started 30 years ago, were made possible through the
Wuppertal-Bonn MRD-CI set of quantum-chemical programs kindly made
available by Prof.\ S.D. Peyerimhoff and Prof.\ R.J. Buenker. The
authors enjoyed many stimulating conversations with Prof.\ Peyerimhoff
on excited electronic states of small molecules over the past decades.
Astrochemistry in Leiden is supported by a Spinoza grant from the
Netherlands Organization for Scientific Research (NWO).

\clearpage

\begin{table}
\begin{center}
\caption{\label{tabc2h}Vertical excitation energies, oscillator
strengths and dominant configurations for C$_2$H at the ground state
equilibrium geometry}
\begin{tabular}{lrrrrcll}
\hline
   & \multicolumn{4}{c}{Energy (eV)} &
$f^{el}$ & Dominant & Type$^a$\\ \cline{2-5}
State & This & Ref. & Ref. & Ref. & This & Configuration  \\
      & work & \cite{Koures91} & \cite{Duflot94} & \cite{Shih77} & work \\
\hline
$1 ^2\Sigma^+$$^b$ &  0.00 & 0.00 & 0.00 & 0.00 & $\ldots$ 
& ...$4\sigma^2 5\sigma^1 1\pi^4 $       & V \\
$2 ^2\Sigma^+$ &  7.06 & 6.73 & 6.63 & 7.32 & 4.0(-4)$^c$ 
& ...$4\sigma^2 5\sigma^1 1\pi^3 2\pi^1$ & V \\
$3 ^2\Sigma^+$ &  8.63 & 8.11 & 8.19 & 9.60 & 4.0(-5) 
 & ...$4\sigma^1 5\sigma^2 1\pi^4 $       & V \\
$4 ^2\Sigma^+$ &  9.28 & 9.01 & 9.21 & 9.18 & 3.0(-3) 
& ...$4\sigma^2 5\sigma^1 1\pi^3 3px^1$  & R \\
$5 ^2\Sigma^+$ &  10.09 & 10.09 &  &  & 2.8(-1) 
& ...$4\sigma^2 5\sigma^1 1\pi^3 4px^1$  & R \\
$6 ^2\Sigma^+$ &  10.28 &  &  &       & 1.0(-5) 
& ...$4\sigma^2 5\sigma^1 1\pi^3 3px^1$  & R \\
$1 ^2\Pi $ &      0.68  & 0.60 & 0.54 & 0.96 & 1.7(-3) 
& ...$4\sigma^2 5\sigma^2 1\pi^3 $       & V \\
$2 ^2\Pi $ &      7.63 & 7.29 & 7.07 & 8.11 & 1.0(-2) 
& ...$4\sigma^2 5\sigma^2 1\pi^2 2\pi^1$ & V \\
$3 ^2\Pi $ &      8.39 & 8.17 & 8.05 & 9.96 & 1.0(-1) 
& ...$4\sigma^2 5\sigma^1 1\pi^3 3s^1$   & R \\
$4 ^2\Pi $ &      9.00 & 8.68 & 8.34 & 8.48 & 3.0(-2) 
& ...$4\sigma^2 5\sigma^1 1\pi^3 3pz^1$  & R \\
$5 ^2\Pi $ &      9.47 &      & 8.70 &  & 3.0(-2) 
& ...$4\sigma^2 5\sigma^1 1\pi^3 3s^1$   & R \\
$6 ^2\Pi $ &      9.96 &      & 8.80 & 9.22 & 2.2(-2) 
& ...$4\sigma^2 5\sigma^1 1\pi^3 4pz^1$  & R \\
$7 ^2\Pi $ &     10.06 &      & 9.25 & 9.78 & 1.1(-2) 
& ...$4\sigma^2 5\sigma^1 1\pi^3 3pz^1$  & R \\
$8 ^2\Pi $ &     10.33 &      & 9.67 & 10.40 & 4.6(-2) 
& ...$4\sigma^2 5\sigma^1 6\sigma^1 1\pi^3$ & M   \\
$1 ^2\Sigma^-$ &  7.61 & 7.48 & 7.34 & 8.13 &  
& ...$4\sigma^2 5\sigma^1 1\pi^3 2\pi^1$ & V \\
$2 ^2\Sigma^-$ &  8.90 & 8.97 & 9.31 & 9.13 &  
& ...$4\sigma^2 5\sigma^1 1\pi^3 2\pi^1$ & V \\
$3 ^2\Sigma^-$ &  9.30 &  &&&
& ...$4\sigma^1 5\sigma^2 1\pi^3 3px^1$  & R \\
$4 ^2\Sigma^-$ &  10.32 & &&&
 & ...$4\sigma^2 5\sigma^1 1\pi^3 3px^1$  & R \\
$5 ^2\Sigma^-$ &  10.46 &  &&&
& ...$4\sigma^2 5\sigma^1 1\pi^3 3px^1$  & R \\
$6 ^2\Sigma^-$ &  10.69 &  &&&
& ...$4\sigma^2 5\sigma^1 1\pi^3 3px^1$  & R \\
$1 ^2\Delta$ &    7.89 & 7.70 & 7.57 & 8.27 &  
& ...$4\sigma^2 5\sigma^1 1\pi^3 2\pi^1$ & V \\
$2 ^2\Delta$ &    8.25 & 8.12 & 7.95 & 8.81 &  
& ...$4\sigma^2 5\sigma^1 1\pi^3 2\pi^1$ & V \\
$3 ^2\Delta$ &    9.20 & 9.06 & 9.23 & 9.12 &  
& ...$4\sigma^2 5\sigma^1 1\pi^3 3px^1$  & R \\
\hline
\end{tabular}
\end{center}
$^a$ V=Valence; R=Rydberg; M=Mixed in this and subsequent tables\\
$^b$ Ground state energy including Davidson correction: -76.432130 Hartree\\
$^c$ Notation x(-y) in this and subsequent tables indicates $x\times 10^{-y}$\\
\end{table}

\clearpage
\begin{table}
\begin{center}
\caption{\label{tabc3}Vertical excitation energies, oscillator
strengths and dominant configurations for C$_3$ at the ground state
equilibrium geometry}
\begin{tabular}{lrrcll}
\hline
   & \multicolumn{2}{c}{Energy (eV)} &
{$f^{el}$} & Dominant & Type \\ \cline{2-3}
State & This & Ref. & This  & Configuration \\
      & work & \cite{Monninger02} & work  \\
\hline
$1 ^1\Sigma_g^+$$^a$ &  0.00 & 0.00 & $\ldots$ 
& ...$4\sigma_g^2 3\sigma_u^2 1\pi_u^4$          & V \\
$2 ^1\Sigma_g^+$ &  5.87 &  &  
& ...$3\sigma_u^2 1\pi_u^4 1\pi_g^2$             & V\\
$3 ^1\Sigma_g^+$ &  8.47 &  &  
& ...$4\sigma_g^2 3\sigma_u^2 1\pi_u^2 1\pi_g^2$ & V \\
$1 ^1\Sigma_u^+$ &  8.17 & 7.97 & 1.1(+0)  
& ...$4\sigma_g^2 3\sigma_u^2 1\pi_u^3 1\pi_g^1$ & V \\
$1 ^1\Pi_u $     &  3.21 & 3.11 & 5.0(-2)  
& ...$4\sigma_g^2 3\sigma_u^1 1\pi_u^4 1\pi_g^1$ & V \\
$2 ^1\Pi_u $ &      8.12 & 7.71 & 7.0(-4) 
& ...$4\sigma_g^1 3\sigma_u^2 1\pi_u^3 1\pi_g^2$ & V \\
$3 ^1\Pi_u $ &     8.64 &  & 7.2(-5) 
& ...$4\sigma_g^1 3\sigma_u^2 1\pi_u^3 1\pi_g^2$ & V \\
$4 ^1\Pi_u $ &    10.13 &  & 2.0(-1) 
& ...$4\sigma_g^2 3\sigma_u^2 1\pi_u^3 3s$       & R \\
$1 ^1\Pi_g $ &      4.01 & 4.00 &  
& ...$4\sigma_g^1 3\sigma_u^2 1\pi_u^4 1\pi_g^1$ & V \\
$2 ^1\Pi_g $ &      7.31 & 7.20  &  
& ...$4\sigma_g^2 3\sigma_u^1 1\pi_u^3 1\pi_g^2$ & V \\
$3 ^1\Pi_g $ &      7.46 & 7.89 &  
& ...$4\sigma_g^2 3\sigma_u^1 1\pi_u^3 1\pi_g^2$ & V \\
$4 ^1\Pi_g $ &      8.10 &  &  
& ...$4\sigma_g^2 3\sigma_u^1 1\pi_u^3 1\pi_g^2$ & V \\
$5 ^1\Pi_g $ &      8.86 &  &  
& ...$4\sigma_g^2 3\sigma_u^1 1\pi_u^3 1\pi_g^2$ & V \\

$1 ^1\Sigma_u^-$ &  4.29 & 3.98 &  
& ...$4\sigma_g^2 3\sigma_u^2 1\pi_u^3 1\pi_g^1$ & V \\
$2 ^1\Sigma_u^-$ &  6.20 &  &  
& ...$4\sigma_g^1 3\sigma_u^1 1\pi_u^4 1\pi_g^2$ & V \\

$1 ^1\Delta_g$ &    5.18 &  &  
& ...$4\sigma_g^2 1\pi_u^4 1\pi_g^2$             & V \\
$2 ^1\Delta_g$ &    9.45 &  &  
& ...$4\sigma_g^2 3\sigma_u^2 1\pi_u^2 1\pi_g^2$ & V \\
$3 ^1\Delta_g$ &    9.78 &  &  
& ...$4\sigma_g^2 3\sigma_u^2 1\pi_u^3 2\pi_u^1$ & V \\
$4 ^1\Delta_g$ &    9.95 &  & 
& ...$3\sigma_u^2 1\pi_u^4 1\pi_g^2$             & V \\
$1 ^1\Delta_u$ &    4.36 & 4.02 &  
& ...$4\sigma_g^2 3\sigma_u^2 1\pi_u^3 1\pi_g^1$ & V \\
\hline
\end{tabular}
\end{center}
$^a$  Ground state energy including Davidson correction: -113.735304 Hartree\\
\end{table}

\clearpage
\begin{table}
\begin{center}
\caption{\label{tabl-c3h}Vertical excitation energies, oscillator
strengths and dominant configurations for $l-$C$_3$H at the ground state
equilibrium geometry}
\begin{tabular}{lrrrcll}
\hline
   & \multicolumn{2}{c}{Energy (eV)} &
{$f^{el}$} & Dominant & Type \\ \cline{2-4}
State & This & Ref. &    This & Configuration  \\
      & work & \cite{Ding01} & work \\
\hline
$1 ^2\Pi $$^a$     &  0.00 & 0.00 & $\ldots$ 
& ...$7\sigma^2 1\pi^4 2\pi^1$       & V \\
$2 ^2\Pi$      &  3.69 & 3.92  & 2.9(-4) 
& ...$7\sigma^2 1\pi^3 2\pi^2$       & V \\
$3 ^2\Pi$      &  4.99 & 5.33  & 1.4(-3) 
& ...$7\sigma^2 1\pi^3 2\pi^2$       & V \\
$4 ^2\Pi$      &  5.39 &  &  7.4(-3)  
& ...$7\sigma^2 1\pi^3 2\pi^2$       & V \\
$5 ^2\Pi$      &  6.51 &  &  6.6(-2)  
& ...$7\sigma^2 1\pi^4 3p_\pi^1$     & R \\
$6 ^2\Pi$      &  7.56 &  &  2.9(-2) 
& ...$7\sigma^1 1\pi^4 2\pi^1 3s$    & R \\
$7 ^2\Pi$      &  7.82 &   & 3.5(-2)  
& ...$7\sigma^2 1\pi^4 3p_\pi^1$     & R \\
$8 ^2\Pi$      &  7.87 &   & 1.3(-1)  
& ...$7\sigma^2 1\pi^4 3p_\pi^1$     & R \\
$1 ^2\Sigma^+$ &  3.45 & 3.66  & 7.0(-3)  
& ...$7\sigma^1 1\pi^4    2\pi^2 $       & V \\
$2 ^2\Sigma^+$ &  5.53 &  &  5.2(-4)  
& ...$7\sigma^2 1\pi^4    3s^1   $       & R \\
$3 ^2\Sigma^+$ &  6.62 &  &  1.5(-2)  
& ...$7\sigma^2 1\pi^4    3p_\sigma^1 $  & R \\
$4 ^2\Sigma^+$ &  7.30 &  &  1.7(-3)  
& ...$7\sigma^1 1\pi^3    2\pi^3 $       & V \\
$5 ^2\Sigma^+$ &  8.15 &  &  2.2(-2)  
& ...$7\sigma^2 1\pi^4    4p_\sigma^1 $  & R \\
$1 ^2\Sigma^-$ &  3.10 & 3.08  & 8.8(-3)  
 & ...$7\sigma^1 1\pi^4    2\pi^2 $       & V \\
$2 ^2\Sigma^-$ &  5.67 &  &  3.5(-3)  
 & ...$7\sigma^1 1\pi^3    2\pi^3 $       & V \\
$3 ^2\Sigma^-$ &  7.15 &  &  5.2(-3)  
& ...$7\sigma^1 1\pi^3    2\pi^3 $       & V \\
$4 ^2\Sigma^-$ &  8.03 &  &  1.1(-2)  
& ...$7\sigma^1 1\pi^4    3\pi^2 $       & R \\
$5 ^2\Sigma^-$ &  8.71 &   & 3.8(-3)  
& ...$7\sigma^1 1\pi^4    2\pi^1 3p_\pi^1 $ & R \\
$1 ^2\Delta$ &    2.79 & 2.96  & 7.6(-3) 
& ...$7\sigma^1 1\pi^4    2\pi^2 $       & V \\
$2 ^2\Delta$ &    6.86 &  &  4.0(-3) 
& ...$7\sigma^1 1\pi^3    2\pi^3 $       & V \\
$3 ^2\Delta$ &    7.63 &  &  2.6(-3) 
& ...$7\sigma^1 1\pi^3    2\pi^3 $       & V \\
$4 ^2\Delta$ &    8.24 &  &  5.8(-3) 
& ...$7\sigma^1 1\pi^4    3\pi^2 $       & R \\
\hline
\end{tabular}
\end{center}
$^a$  Ground state energy including Davidson correction: -114.381395 Hartree\\
\end{table}

\clearpage
\begin{table}
\begin{center}
\caption{\label{tabc-c3h}Vertical excitation energies, oscillator
strengths and dominant configurations for $c-$C$_3$H at the ground state
equilibrium geometry}
\begin{tabular}{lrrcll}
\hline
   & \multicolumn{2}{c}{Energy (eV)} &
{$f^{el}$} & Dominant & Type \\ \cline{2-3}
State & This & Ref. & This  & Configuration  \\
      & work & \cite{Ding01} & work \\
\hline
$1 ^2B_2$$^a$ &  0.00 & 0.00 & $\ldots$ 
& $...5a_1^2 6a_1^2 1b_1^2 2b_2^2 3b_2^1$    & V \\
$2 ^2B_2$ &  6.79 & 6.62 & 4.0(-3) 
 & $2b_2 \to 3b_2$                              & V \\
$3 ^2B_2$ &  6.94 &  & 6.4(-3) 
& $6a_1 \to 7a_1$                              & R \\
$4 ^2B_2$ &  7.54 &  & 8.7(-3) 
& $3b_2 \to 4b_2$                              & R \\
$5 ^2B_2$ &  7.80 &  & 5.3(-3) 
 & $6a_1 \to 8a_1$                              & R \\
$6 ^2B_2$ &  7.81 &  & 2.7(-4) 
 & $1b_1 \to 2b_1$                              & V \\
$7 ^2B_2$ &  8.74 &  & 1.3(-3)
& $6a_1 \to 7a_1$                              & R \\
$8 ^2B_2$ &  8.89 &  & 5.6(-3) 
 & $3b_2 \to 5b_2$                              & R \\
$9 ^2B_2$ &  9.43 &  & 1.5(-3) 
& $6a_1 \to 8a_1$                              & R \\
$1 ^2A_1$ &  1.22 & 1.22  & 1.7(-2) 
& $6a_1 \to 3b_2$                              & V \\
$2 ^2A_1$ &  5.06 & 5.17 & 3.1(-2) 
& $5a_1 \to 3b_2$                              & V \\
$3 ^2A_1$ &  6.76 &  & 1.9(-5) 
& $3b_2 \to 7a_1$                              & R \\
$4 ^2A_1$ &  7.37 &  & 9.3(-3) 
& $3b_2 \to 8a_1$                              & R \\
$5 ^2A_1$ &  8.46 &  & 3.7(-3) 
& $3b_2 \to 9a_1$                              & R \\
$6 ^2A_1$ &  8.73 &  & 8.0(-3) 
& $1b_1 \to 2a_2$                              & V \\
$7 ^2A_1$ &  8.82 &  & 4.9(-3) 
& $3b_2 \to 10a_1$                              & R \\
$8 ^2A_1$ &  9.14 &  & 1.1(-5) 
 & $6a_1 1b_1 \to 4b_1 3b_2$                    & V \\
$9 ^2A_1$ &  9.97 &  & 6.9(-4) 
& $(6a_1)^2  \to 7a_1 3b_2$                    & R \\
$1 ^2A_2$ &  3.90 & 3.94 & 2.2(-3)  
& $3b_2 \to 2a_2$                              & V \\
$2 ^2A_2$ &  4.61 & 4.94 & 1.8(-2)  
& $6a_1 \to 2b_1$                              & V \\
$3 ^2A_2$ &  5.53 &  & 4.2(-4)  
& $6a_1 \to 3b_1$                              & V \\
$4 ^2A_2$ &  7.50 &  & 2.7(-2)  
& $(6a_1)^2 \to 3b_2 2a_2$                     & R \\
$5 ^2A_2$ &  7.54 &  & 1.8(-2)  
& $6a_1 \to 4b_1$                              & R\\
$6 ^2A_2$ &  8.53 &  & 1.3(-2)  
& $5a_1 \to 2b_1$                              & R \\
$7 ^2A_2$ &  8.72 &  & 9.6(-3)  
 & $3b_2 \to 1a_2$                              & R \\
$8 ^2A_2$ &  8.99 &  & 7.3(-3)  
& $6a_1 \to 3b_1$                              & R \\
$9 ^2A_2$ &  9.46 &  & 4.0(-3)  
 & $6a_1 \to 4b_1$                              & R \\
$1 ^2B_1$ &  3.47 & 3.49 &   
& $3b_2 \to 2b_1$                              & V \\
$2 ^2B_1$ &  4.70 & 4.63 &   
& $3b_2 \to 3b_1$                              & V \\
$3 ^2B_1$ &  5.73 &  &   
& $6a_1 \to 1a_2$                              & V \\
$4 ^2B_1$ &  5.97 &  &   
& $6a_1 \to 2a_2$                              & V \\
$5 ^2B_1$ &  7.19 &  &   
& $(6a_1)^2 \to 2b_1 3b_2$                     & V \\
$6 ^2B_1$ &  7.51 &  &   
& $3b_2 \to 4b_1$                              & R\\
$7 ^2B_1$ &  8.80 &  &  
& $3b_2 \to 2b_1$                              & R \\
$8 ^2B_1$ &  8.97 &  &   
& $6a_1 \to 1a_2$                              & R \\
$9 ^2B_1$ &  9.09 &  &   
& $5a_1 \to 2a_2$                              & R \\
\hline
\end{tabular}
\end{center}
$^a$  Ground state energy including Davidson correction: -114.393014 Hartree\\
$^a$ footnote \\
\end{table}

\clearpage
\begin{table}
\begin{center}
\caption{\label{tabl-c3h2}Vertical excitation energies, oscillator
strengths and dominant configurations for HC$_3$H at the ground state
equilibrium geometry}
\begin{tabular}{lrcll}
\hline
   & Energy (eV) &
{$f^{el}$} & Dominant & Type \\ 
State & This & This  & Configuration  \\
      & work & work \\
\hline
$1 ^3B$$^a$ &  0.00 & $\ldots$ 
& ...$6a^2 7a^1 3b^2 4b^1$                     & V \\
$2 ^3B$ &  4.37 & 1.8(-4) 
& $3b \to 4b$                                    & V \\
$3 ^3B$ &  4.58 & 1.2(-5) 
& $6a \to 7a$                                    & V \\
$4 ^3B$ &  5.61 & 4.2(-5) 
& $7a \to 9a$                                    & R \\
$5 ^3B$ &  5.89 & 7.5(-4) 
& $7a \to 8a$                                    & V \\
$6 ^3B$ &  6.08 & 6.7(-4) 
 & $4b \to 5b$                                    & V \\
$7 ^3B$ &  6.67 & 7.0(-6) 
& $7a \to 10a$                                   & R \\
$8 ^3B$ &  6.93 & 6.8(-4) 
& $4b \to 6b$                                    & R \\
$9 ^3B$ &  6.97 & 1.9(-4) 
& $7a \to 11a$                                   & R \\
$1 ^3A$ &  4.23 & 1.7(-2) 
 & $7a \to 4b$                                    & V \\
$2 ^3A$ &  4.33 & 2.4(-4) 
 & $3b \to 7a$                                    & V \\
$3 ^3A$ &  5.66 & 9.3(-4) 
& $4b \to 8a$                                    & V \\
$4 ^3A$ &  5.97 & 1.7(-3) 
& $7a \to 5b$                                    & V\\
$5 ^3A$ &  6.29 & 1.5(-1) 
& $7a \to 6b$                                    & R \\
$6 ^3A$ &  6.68 & 1.2(-3) 
& $4b \to 9a$                                    & R \\
$7 ^3A$ &  6.95 & 4.8(-3) 
& $4b \to 10a$                                   & R \\
$8 ^3A$ &  7.52 & 2.4(-1) 
& $4b \to 12a$                                   & R \\
$9 ^3A$ &  7.73 & 2.2(-3) 
 & $4b \to 13a$                                   & R \\
\hline
\end{tabular}
\end{center}
$^a$  Ground state energy including Davidson correction: -114.924751 Hartree\\
\end{table}

\clearpage
\begin{table}
\begin{center}
\caption{\label{tabc-c3h2}Vertical excitation energies, oscillator
strengths and dominant configurations for $c-$C$_3$H$_2$
at the ground state equilibrium geometry}
\begin{tabular}{lrcll}
\hline
      & Energy (eV) & {$f^{el}$} & Dominant & Type\\ 
State & This & This  & Configuration  \\
      & work & work \\
\hline
$1 ^1A_1$$^a$ &  0.00 & $\ldots$ 
& $...5a_1^2 6a_1^2 1b_1^2 2b_2^2 3b_2^2$    & V \\
$2 ^1A_1$ &  6.31 & 1.1(-4) 
& $6a_1 \to 8a_1$                              & R \\
$3 ^1A_1$ &  6.85 & 6.7(-2) 
& $6a_1 \to 9a_1$                              & R \\
$4 ^1A_1$ &  7.82 & 1.1(-2) 
& $6a_1 \to 10a_1$                             & R \\
$5 ^1A_1$ &  8.11 & 2.9(-3) 
& $6a_1 \to 9a_1$                              & R \\
$6 ^1A_1$ &  8.57 &  7.0(-2)
 & $6a_1 \to 11a_1$                             & R \\
$7 ^1A_1$ &  9.37 &  1.2(-1)
& $1b_1 \to 2b_1$                              & V \\
$8 ^1A_1$ &  9.70 &  4.8(-2)
& $6a_1 \to 12a_1$                             & R \\
$9 ^1A_1$ &  10.94 &  8.7(-4)
& $6a_1 \to 12a_1$                             & R \\
$1 ^1B_1$ &  4.89 &  2.8(-2)
& $6a_1 \to 5b_1$                              & V \\
$2 ^1B_1$ &  6.65 &  2.4(-3)
& $3b_2 \to 1a_2$                              & V \\
$3 ^1B_1$ &  6.91 &  1.9(-2)
& $6a_1 \to 2b_1$                              & R \\
$4 ^1B_1$ &  8.25 &  6.3(-3)
& $6a_1 \to 3b_1$                              & R \\
$5 ^1B_1$ &  8.63 &  1.1(-2)
& $1b_1 \to 8a_1$                              & R \\
$6 ^1B_1$ &  10.30 &  4.3(-2)
 & $1b_1 \to 9a_1$                              & R \\
$7 ^1B_1$ &  11.16&  2.6(-2)
& $3b_2 \to 1a_2$                              & R \\
$8 ^1B_1$ &  11.53&  1.8(-4)
 & $1b_1 \to 10a_1$                             & R \\
$9 ^1B_1$ &  12.30 &  1.3(-3)
 & $6a_1 \to 4b_1$                              & R \\
$1 ^1B_2$ &  6.68 &  1.4(-4)
& $6a_1 \to 6b_2$                              & R \\
$2 ^1B_2$ &  8.13 &  4.4(-3)
& $6a_1 \to 7b_2$                              & R \\
$3 ^1B_2$ &  8.29 &  1.1(-2)
& $6a_1 \to 4b_2$                              & V \\
$4 ^1B_2$ &  9.01 &  6.6(-2) 
& $1b_1 \to 1a_2$                              & R \\
$5 ^1B_2$ &  9.20 &  5.8(-3)
& $3b_2 \to 8a_1$                              & R \\
$6 ^1B_2$ &  9.99 &  3.0(-2)
 & $3b_2 \to 9a_1$                              & R \\
$7 ^1B_2$ & 10.58 &  3.0(-1)
& $6a_1 \to 5b_2$                              & V \\
$8 ^1B_2$ & 11.03 & 1.6(-5) 
 & $3b_2 \to 8a_1$                              & R \\
$9 ^1B_2$ & 11.23 &  1.0(-1)
& $6a_1 \to 8b_2$                              & R \\
$1 ^1A_2$ &  3.92 &  
& $6a_1 \to 1a_2$                              & V \\
$2 ^1A_2$ &  7.03 & 
& $3b_2 \to 5b_1$                              & V \\
$3 ^1A_2$ &  8.46 &  
& $6a_1 \to 2a_2$                              & R \\
$4 ^1A_2$ &  9.25 &  
& $5a_1 \to 2a_2$                              & V \\
$5 ^1A_2$ &  9.94 &  
 & $3b_2 \to 2b_1$                              & R \\
$6 ^1A_2$ &  10.86 &  
 & $1b_1 \to 4b_2$                              & R \\
$7 ^1A_2$ &  11.28 &  
 & $2b_2 \to 3b_1$                              & R \\
$8 ^1A_2$ &  12.31 &  
& $1b_1 \to 4b_2$                              & R \\
$9 ^1A_2$ &  12.65&  
& $2b_2 \to 3b_1$                              & R \\
\hline
\end{tabular}
\end{center}
$^a$  Ground state energy including Davidson correction: -115.037418 Hartree\\
\end{table}

\clearpage
\begin{table}
\begin{center}
\caption{\label{tabv-c3h2}Vertical excitation energies, oscillator
strengths and dominant configurations for $l-$C$_3$H$_2$
at the ground state equilibrium geometry}
\begin{tabular}{lrcll}
\hline
      & Energy (eV) & {$f^{el}$} & Dominant & Type\\ 
State & This & This  & Configuration   \\
      & work & work \\
\hline
$1 ^1A_1$$^a$ &  0.00 & $\ldots$ 
& $...6a_1^2 7a_1^2 1b_1^2 2b_2^2$           & V \\
$2 ^1A_1$ &  5.54 & 1.3(-1) 
& $1b_1 \to 2b_1$                              & V \\
$3 ^1A_1$ &  5.98 & 4.7(-2) 
& $(2b_2)^2 \to (2b_1)^2$                      & V \\
$4 ^1A_1$ &  7.79 & 9.8(-2) 
& $2b_2 \to 3b_2$                              & R \\
$5 ^1A_1$ &  8.87 & 1.5(-2) 
& $2b_2 \to 4b_2$                              & R \\
$6 ^1A_1$ &  9.00 & 3.0(-1) 
& $2b_2 \to 5b_2$                              & R \\
$7 ^1A_1$ &  9.38 &  3.0(-1)
& $2b_2 \to 6b_2$                              & R \\
$8 ^1A_1$ &  9.84 &  8.8(-2)
& $2b_2 \to 7b_2$                              & R \\
$9 ^1A_1$ &  10.02 &  6.2(-2)
& $1b_1 \to 3b_1$                              & R \\
$1 ^1B_1$ &  2.44 &  8.8(-3)
& $7a_1 \to 2b_1$                              & V \\
$2 ^1B_1$ &  6.36 &  1.3(-2)
 & $7a_1 1b_1 \to (2b_1)^2$                     & V \\
$3 ^1B_1$ &  8.01 &  1.7(-2)
 & $7a_1 \to 3b_1$                              & R \\
$4 ^1B_1$ &  8.82 &  1.3(-3)
& $7a_1 2b_2 \to 2b_1 5b_2$                    & V \\
$5 ^1B_1$ &  8.86 &  3.8(-3)
 & $7a_1 \to 4b_1$                              & R \\
$6 ^1B_1$ &  8.88 & 7.9(-2) 
 & $1b_1 \to 10a_1$                             & R \\
$7 ^1B_1$ &  9.11 &  2.1(-2)
 & $7a_1 \to 5b_1$                              & R \\
$8 ^1B_1$ &  9.85 &  3.6(-3)
& $7a_1 \to 6b_1$                              & R \\
$9 ^1B_1$ &  10.13 &  1.9(-2)
 & $1b_1 \to 8a_1$                              & R \\
$1 ^1B_2$ &  6.06 &  2.4(-4)
& $6a_1 2b_2 \to (2b_1)^2$                     & V \\
$2 ^1B_2$ &  6.92 &  1.3(-2)
& $2b_2 \to 8a_1$                              & V \\
$3 ^1B_2$ &  8.07 &  4.0(-2)
 & $2b_2 \to 9a_1$                              & R \\
$4 ^1B_2$ &  8.87 &  4.2(-4) 
& $2b_2 \to 10a_1$                             & R \\
$5 ^1B_2$ &  8.98 &  2.0(-4)
& $2b_2 \to 11a_1$                             & R \\
$6 ^1B_2$ &  9.36 &  1.9(-2)
& $2b_2 \to 12a_1$                             & R \\
$7 ^1B_2$ &  9.68 &  7.2(-5)
& $2b_2 \to 13a_1$                             & R\\
$8 ^1B_2$ &  10.14 &  3.4(-2)
& $2b_2 \to 14a_1$                             & R \\
$9 ^1B_2$ &  10.50&  $<$1(-6)
& $2b_2 \to 15a_1$                             & R \\
$1 ^1A_2$ &  1.86 &  
& $2b_2 \to 2b_1$                              & V \\
$2 ^1A_2$ &  6.68 &  
& $1b_1 2b_2 \to (2b_1)^2$                     & V \\
$3 ^1A_2$ &  7.71 &  
& $2b_2 \to 3b_1$                              & R \\
$4 ^1A_2$ &  8.30 &  
 & $2b_2 \to 4b_1$                              & V \\
$5 ^1A_2$ &  8.86 &  
 & $2b_2 \to 5b_1$                              & R \\
$6 ^1A_2$ &  8.88 &  
 & $1b_1 \to 3b_2$                              & V \\
$7 ^1A_2$ &  9.00 &  
& $2b_2 \to 6b_1$                              & R\\
$8 ^1A_2$ &  9.17 &  
 & $2b_2 \to 7b_1$                              & R \\
$9 ^1A_2$ &  9.91 &  
& $2b_2 \to 8b_1$                              & R \\
\hline
\end{tabular}
\end{center}
$^a$  Ground state energy including Davidson correction: -114.965558 Hartree\\
\end{table}

\clearpage
\begin{table}
\begin{center}
\caption{\label{tabc4}Vertical excitation energies and oscillator
strengths and dominant configurations for $l-$C$_4$
at the ground state equilibrium geometry}
\begin{tabular}{lrrrcll}
\hline
   & \multicolumn{3}{c}{Energy (eV)} & $f^{el}$ & Dominant & Type\\ \cline{2-4}
State & This & Ref.                & Exp.         & This  & Configuration \\
      & work & \cite{Masso06}  &\cite{Masso06} & work \\
\hline
$1 ^3\Sigma_g^-$$^a$ &  0.00 & 0.00 & 0.00 & $\ldots$ 
& ...$4\sigma_u^2 5\sigma_g^2 1\pi_u^4 1\pi_g^2$  & V \\
$2 ^3\Sigma_g^-$ &  6.40 & 6.11 & &  
& ...$4\sigma_u^2 5\sigma_g^2 1\pi_u^3 1\pi_g^3$  & V \\
$3 ^3\Sigma_g^-$ &  6.79 &  & &  
 & ...$4\sigma_u^2 5\sigma_g^2 1\pi_u^3 1\pi_g^2 2\pi_u^1$ & V \\
$4 ^3\Sigma_g^-$ &  7.05 &  & &   
& ...$4\sigma_u^2 5\sigma_g^2 1\pi_u^3 1\pi_g^2 2\pi_u^1$ & V \\
$1 ^3\Sigma_g^+$ &  5.80 &  & &   
& ...$4\sigma_u^1 5\sigma_g^1 1\pi_u^4 1\pi_g^3 2\pi_u^1$ & V \\
$1 ^3\Sigma_u^-$ &  3.61 & 3.74 & 3.27 & 9.4(-03)  
& ...$4\sigma_u^2 5\sigma_g^2 1\pi_u^3 1\pi_g^3$  & V \\
$2 ^3\Sigma_u^-$ &  6.95 &  & &1.6(+0) 
 & ...$4\sigma_u^2 5\sigma_g^2 1\pi_u^4 1\pi_g^1 2\pi_u^1$  & V \\
$1 ^3\Sigma_u^+$ &  1.68 &  & &  
 & ...$4\sigma_u^1 5\sigma_g^1 1\pi_u^4 1\pi_g^4$          & V \\
$2 ^3\Sigma_u^+$ &  2.82 & 2.84 &&  
 & ...$4\sigma_u^2 5\sigma_g^2 1\pi_u^3 1\pi_g^3$          & V \\
$3 ^3\Sigma_u^+$ &  4.30 &  &&   
& ...$4\sigma_u^2 5\sigma_g^2 1\pi_u^4 1\pi_g^1 2\pi_u^1$ & V \\
$4 ^3\Sigma_u^+$ &  4.65 &  &&   
& ...$4\sigma_u^2 1\pi_u^4 1\pi_g^3 2\pi_u^1$             & V \\
$5 ^3\Sigma_u^+$ &  5.63 &  &&   
 & ...$5\sigma_g^2 1\pi_u^4 1\pi_g^3 2\pi_u^1$ & V \\
$1 ^3\Pi_g $ &      0.95 & 1.09 & 0.82 &  
 & ...$4\sigma_u^2 5\sigma_g^1 1\pi_u^4 1\pi_g^3$  & V \\
$2 ^3\Pi_g $ &      4.26 &   &&  
 & ...$4\sigma_u^1 5\sigma_g^2 1\pi_u^3 1\pi_g^4$  & V \\
$3 ^3\Pi_g $ &      5.31 &  &&  
 & ...$4\sigma_u^1 5\sigma_g^2 1\pi_u^4 1\pi_g^2 2\pi_u^1$  & V \\
$4 ^3\Pi_g $ &      5.96 &  &&  
& ...$4\sigma_u^1 5\sigma_g^2 1\pi_u^4 1\pi_g^2 2\pi_u^1$  & V \\
$1 ^3\Pi_u $     &  1.19 & 1.37 & 0.93 & 4.6(-3) 
 & ...$4\sigma_u^1 5\sigma_g^2 1\pi_u^4 1\pi_g^3$  & V \\
$2 ^3\Pi_u $ &      4.00 &  && 2.2(-4) 
& ...$4\sigma_u^2 5\sigma_g^1 1\pi_u^3 1\pi_g^4$  & V \\
$3 ^3\Pi_u $ &     5.10 &  && 1.6(-2) 
& ...$4\sigma_u^2 5\sigma_g^1 1\pi_u^4 1\pi_g^2 2\pi_u^1$  & V \\
$4 ^3\Pi_u $ &   5.70 &  && 2.8(-5) 
& ...$4\sigma_u^2 5\sigma_g^1 1\pi_u^4 1\pi_g^2 2\pi_u^1$  & V \\
$5 ^3\Pi_u $ &     5.86 &  && 1.2(-1) 
& ...$4\sigma_u^2 5\sigma_g^1 1\pi_u^4 1\pi_g^2 2\pi_u^1$  & V \\
$1 ^3\Delta_u$ &  2.77 & 2.82 &&  
 & ...$4\sigma_u^2 5\sigma_g^2 1\pi_u^3 1\pi_g^3$  & V \\
$2 ^3\Delta_u$ &  4.25 & 3.56 &&  
 & ...$4\sigma_u^2 5\sigma_g^2 1\pi_u^4 1\pi_g^1 2\pi_u^1$  & V \\
$1 ^3\Delta_g$  &  6.67 &  &&  
 & ...$4\sigma_u^2 5\sigma_g^2 1\pi_u^3 1\pi_g^2 2\pi_u^1$ & V \\
\hline
\end{tabular}
\end{center}
$^a$  Ground state energy including Davidson correction: -152.228684 Hartree\\
\end{table}

\clearpage
\begin{table}
\begin{center}
\caption{\label{tabc4h}Vertical excitation energies,oscillator
strengths and dominant configurations for $l-$C$_4$H
at the ground state equilibrium geometry}
\begin{tabular}{lrrccll}
\hline
   & \multicolumn{2}{c}{Energy (eV)} &
\multicolumn{2}{c}{$f^{el}$} & Dominant & Type \\ \cline{2-5}
State & This & Ref.             & This & Ref.  & Configuration \\
      & work & \cite{Graf01} & work & \cite{Graf01}\\
\hline
$1 ^2\Sigma^+$$^a$ &  0.00 & 0.00 & $\ldots$ & $\ldots$
 & ...$9\sigma^1 1\pi^4    2\pi^4 $       & V \\
$2 ^2\Sigma^+$ &  5.11 & 4.76 & 1.0(-6) & $<$7(-7)
& ...$9\sigma^1 1\pi^4    2\pi^3  3\pi^1$& V \\
$3 ^2\Sigma^+$ &  7.36 & 6.62 & 9.5(-4) & 1.8(-3)
& ...$9\sigma^1 1\pi^3    2\pi^4  3\pi^1$& V \\
$4 ^2\Sigma^+$ &  7.74 &      & 3.5(-1) & 
 & ...$9\sigma^1 1\pi^4    3\pi^3  3p_\pi^1$ & R \\
$5 ^2\Sigma^+$ &  8.76 &      & 5.5(-1) & 
 & ...$9\sigma^1 1\pi^4    2\pi^3  3p_\pi^1$ & R \\
$1 ^2\Pi$      &  0.37 & 0.44  & 7.8(-4) & 8.4(-4)
 & ...$9\sigma^2 1\pi^4 2\pi^3$       & V \\
$2 ^2\Pi$      &  3.59 & 3.31 & 7.8(-4) & 9.8(-4)
& ...$9\sigma^2 1\pi^3 2\pi^4$       & V \\
$3 ^2\Pi$      &  5.31 & 4.71 & 3.4(-4) & 7.2(-4)
& ...$9\sigma^2 1\pi^4 2\pi^3 3\pi^1$ & V \\
$4 ^2\Pi$      &  7.15 & 5.92 & 2.2(-3) & 9.2(-4)
 & ...$9\sigma^1 10\sigma^1 1\pi^4 2\pi^3$  & R \\
$5 ^2\Pi$      &  7.85 & 6.82 & 7.0(-3) & 8.8(-5)  
 & ...$9\sigma^1 10\sigma^1 1\pi^4 2\pi^3$  & R \\
$6 ^2\Pi$      &  7.90 & 7.84 & 1.4(-2)  &
 & ...$9\sigma^1 10\sigma^1 1\pi^4 2\pi^3$  & R \\
$7 ^2\Pi$      &  8.41 &      & 2.2(-3)  &
 & ...$9\sigma^1 11\sigma^1 1\pi^4 2\pi^3$  & R \\
$1 ^2\Sigma^-$ &  6.03 &      &  & 
 & ...$9\sigma^1 1\pi^4    2\pi^3  3\pi^1$& V \\
$2 ^2\Sigma^-$ &  6.48 &      &  & 
& ...$9\sigma^1 1\pi^4    2\pi^3  3\pi^1$& V \\
$3 ^2\Sigma^-$ &  8.58 &      &  & 
& ...$9\sigma^1 1\pi^4    2\pi^3  3p_\pi^1$& R \\
$4 ^2\Sigma^-$ &  8.86 &      &  & 
 & ...$9\sigma^1 1\pi^4    2\pi^3  3p_\pi^1$& R \\
$1 ^2\Delta$   &  5.76 & 5.20 &  & 
& ...$9\sigma^1 1\pi^4  2\pi^3 3\pi^1 $       & V \\
$2 ^2\Delta$   &  5.97 & 5.21 &  & 
 & ...$9\sigma^1 1\pi^4  2\pi^3 3\pi^1 $       & V \\
$3 ^2\Delta$   &  7.85 &      &  & 
 & ...$9\sigma^1 1\pi^3  2\pi^3 3p_\pi^1$      & R \\
$4 ^2\Delta$   &  8.10 &      &  & 
 & ...$9\sigma^1 1\pi^3  2\pi^3 3p_\pi^1$      & R \\
\hline
\end{tabular}
\end{center}
$^a$  Ground state energy including Davidson correction: -152.228684 Hartree\\
\end{table}

\clearpage
\begin{table}
\begin{center}
\caption{\label{tabc5h}Vertical excitation energies, oscillator
strengths and dominant configurations for $l-$C$_5$H
at the ground state equilibrium geometry} 
\begin{tabular}{lrrccll}
\hline
   & \multicolumn{2}{c}{Energy (eV)} &
\multicolumn{2}{c}{$f^{el}$} & Dominant & Type \\ \cline{2-5}
State & This & Ref.             & This & Ref.  & Configuration \\
      & work & \cite{Haubrich02}$^b$ & work & \cite{Haubrich02}$^b$\\
\hline
$1 ^2\Pi$$^a$      &  0.00 & 0.00  & $\ldots$ & $\ldots$
& ...$11\sigma^2 1\pi^4 2\pi^4 3\pi^1$          & V \\
$2 ^2\Pi$      &  3.05 & 3.21 & 5.1(-4) & 1(-3) 
& ...$11\sigma^2 1\pi^4 2\pi^3 3\pi^2$          & V \\
$3 ^2\Pi$      &  3.91 & 3.99 & 1.8(-3) & 4(-3) 
 & ...$11\sigma^2 1\pi^4 2\pi^3 3\pi^2$          & V \\
$4 ^2\Pi$      &  4.18 & 4.19 & 5.7(-2) & 3(-2) 
 & ...$11\sigma^2 1\pi^4 2\pi^3 3\pi^2$          & V \\
$5 ^2\Pi$      &  5.20 & 5.10 & 3.5(-3) & 5(-3)
& ...$11\sigma^2 1\pi^3 2\pi^4 3\pi^2$          & V \\
$6 ^2\Pi$      &  6.09 & 6.35 & 5.2(-2) & 1(-3) 
& ...$11\sigma^2 1\pi^3 2\pi^4 3\pi^2$          & V \\
$7 ^2\Pi$      &  6.13 & 6.11 & 3.0(-3) & 1.4(-1)
 & ...$11\sigma^2 1\pi^3 2\pi^4 3\pi^2$          & V \\
$1 ^2\Sigma^+$ &  3.21 & 3.19 & 4.6(-3) & 9(-3)
& ...$11\sigma^1 1\pi^4 2\pi^4   3\pi^2 $       & V \\
$2 ^2\Sigma^+$ &  5.07 & 6.28 & 3.4(-3) & 1(-3) 
 & ...$11\sigma^2 1\pi^4 2\pi^4   3s^1   $       & R \\
$3 ^2\Sigma^+$ &  5.73 & 7.01 &  1.0(-6)& 2(-3)
 & ...$11\sigma^2 1\pi^4 2\pi^4   3p_\sigma^1$   & R \\
$4 ^2\Sigma^+$ &  6.69 &  &  1.9(-3)& 
& ...$11\sigma^2 1\pi^4 2\pi^4   4s^1 $         & R \\
$5 ^2\Sigma^+$ &  7.09  &  & 2.0(-6) & 
 & ...$11\sigma^2 1\pi^3 2\pi^3   3\pi^2$        & V \\
$1 ^2\Sigma^-$ &  2.83 & 2.80  & 6.0(-3) & 7(-3)
& ...$11\sigma^1 1\pi^4 2\pi^4   3\pi^2 $       & V \\
$2 ^2\Sigma^-$ &  5.12 & 5.25 & 1.6(-3) & 1(-3)
 & ...$11\sigma^1 1\pi^4 2\pi^3   3\pi^3 $       & V \\
$3 ^2\Sigma^-$ &  5.99 & 6.22   &  2.6(-3)& 1(-3)
& ...$11\sigma^1 1\pi^4 2\pi^3   3\pi^3 $       & V \\
$1 ^2\Delta$   &  2.71 & 2.62 & 4.8(-3)  & 3(-3)
 & ...$11\sigma^1 1\pi^4 2\pi^4   3\pi^2 $       & V \\
$2 ^2\Delta$   &  6.06 & 5.92 & 2.6(-3) & 1(-3) 
 & ...$11\sigma^1 1\pi^4 2\pi^3   3\pi^3 $       & V \\
$3 ^2\Delta$   &  6.63 & 6.05  & 1.5(-3) & 2(-3)
 & ...$11\sigma^2 12\sigma^1 1\pi^4 2\pi^4$       & V \\
\hline
\end{tabular}
\end{center}
$^a$ Ground state energy including Davidson correction: -189.983327 Hartree \\
$^b$ Their cc-p-VTZ+SP triple $\zeta$ results if available, double $\zeta$
results otherwise.
\end{table}

\clearpage
\begin{table}
\begin{center}
\caption{\label{tabdiss} Dissociation and ionization energies$^a$ 
with respect to
ground-state equilibrium energy (in eV) for various species}
\begin{tabular}{llrrrr}
\hline 
Species & Products & $\Delta E_{\rm diss}$ & Ref. &
$\Delta E_{\rm ion}$ & Ref.\\ 
\hline 
$l-$C$_3$& C$_2$ + C($^3P$) & 4.63 & \cite{Kim97} & 12.1 & \cite{Benedikt05} \\
$l-$C$_4$ & C$_3$ + C($^3P$) & 4.71 & \cite{Choi00} & 10.7 & \cite{Ortiz93} \\
$l-$C$_2$H & C$_2$ + H & 4.90 & TW$^b$  & 11.4 & TW\\ 
$l-$C$_3$H & C$_3$ + H & 3.27 & TW & 8.6 & TW\\ 
$c-$C$_3$H & C$_3$ + H & 4.29 & TW & 9.6 & TW \\ 
$l-$C$_4$H & C$_4$ + H & 4.65 & TW & 9.6 & TW \\ 
$l-$C$_5$H & C$_5$ + H & 3.56 & TW & 7.4 & TW \\
HC$_3$H & C$_3$H + H & 3.1 & \cite{Mebel98} 
   &8.96 & \cite{Taatjes05} \\ 
$c-$C$_3$H$_2$ & C$_3$H + H & 4.37 & \cite{Mebel98} & 9.15&\cite{Clauberg92}\\ 
$l-$C$_3$H$_2$ & C$_3$H + H & 3.87 & \cite{Mebel98}& 10.43&\cite{Clauberg92}\\ 
\hline
\end{tabular}
\end{center}
\smallskip
$^a$ Experimental data refer to the adiabatic ionization potentials; computed
values in this work to vertical ionization potentials \\
$^b$ Computed in this work, using the same basis set and procedure for the
products; dissociation energies were obtained at 20 Bohr \\
\end{table}

\clearpage
\begin{table}
\begin{center}
\caption{\label{tabisrates} Photodissociation rates (in s$^{-1}$) for
various molecules in the unshielded interstellar radiation field}
\begin{tabular}{lr}
\hline 
Species & Rate \\
\hline 
$l-$C$_3$      & 5.0(-9) \\ 
$l-$C$_4$      & 8.5(-9) \\
$l-$C$_2$H     & 1.6(-9) \\
$l-$C$_3$H     & 1.8(-9) \\
$c-$C$_3$H     & 1.1(-9) \\
$l-$C$_4$H     & 3.7(-9) \\
$l-$C$_5$H     & 1.3(-9) \\
HC$_3$H        & 2.2(-9) \\
$c-$C$_3$H$_2$ & 1.4(-9) \\
$l-$C$_3$H$_2$ & 5.1(-9) \\ 
\hline
\end{tabular}
\end{center}
\end{table}

 

\end{document}